\documentstyle[preprint,revtex]{aps}

\begin{document}

\preprint{IFUSP/P-1031}
\preprint{gr-qc/9306026}

\draft

\begin{title}
Gauge fields on Riemann-Cartan space-times
\end{title}

\author{Alberto Saa}
\begin{instit}
Instituto de F\'{\i}sica,\\
Universidade de S\~ao Paulo, Caixa Postal 20516 \\
01498-970 S\~ao Paulo, S.P.\\
Brazil
\end{instit}

\begin{abstract}
Gauge fields are described on an Riemann-Cartan space-time by means of
tensor-valued
differential forms and exterior calculus. It is shown that minimal coupling
procedure leads to a gauge invariant theory where gauge fields
interact with torsion, and that consistency conditions for the gauge fields
impose restrictions in the non-Riemannian structure of space-time.
The new results differ from the well established ones obtained by using
minimal coupling procedure at the action formulation. The sources of
these differences are pointed out and discussed.
\end{abstract}

\newpage

\section{Introduction}

The Einstein-Cartan theory is the natural theory of gravity that emerges
from the local gauge theory for the Poincar\'e group, and it is
in accordance with the present day experimental data\cite{hehl,venzo}.
This theory
has been discussed in recent years, and in particular the problem of
coupling gauge fields to Riemann-Cartan space-time $U_4$ has been studied
(see for example \cite{odintsov} and references therein).
The wide spread conclusion that gauge fields don't couple minimally to the
non-Riemannian structure of space-time arises from an analysis using minimal
coupling procedure (MCP) at the Action level.

In this work it is shown that, by using MCP in the Minkowskian
equations of motion written by means of tensor-valued differential
forms, one gets the
$U_4$ equations of motion which will allow the interaction between
gauge fields and the torsion.
In order to have consistent equations we are lead to
the restriction that the trace of the torsion tensor must be derived from a
scalar potential. With this condition it is possible to get the $U_4$
equations of motion by using MCP at the Action level, provided that we
introduce the invariant and  covariantly constant $U_4$ volume
 element \cite{saa}.

The work is organized in 5 sections, where the first is this introduction.
In section 2, basic
facts on Riemann-Cartan geometry are briefly presented. Maxwell
fields are described on an $U_4$ manifold in Section 3. In  Section 4,
the results of Section 3 are generalized to the non-abelian case. In the
last section, the dynamics for the $U_4$ geometry and the similarity
between the new results and the dilaton gravity are discussed. Yet in the
last section, it is
shown that the problems with gauge fields on $U_4$ are
connected with the Hodge star operator $({}^*)$, which in $U_4$ space-time
must be different from the usual one of $V_4$ space-time.

\newpage

\section{The $U_4$ Manifold}

The Riemann-Cartan space-time $U_4$ is characterized by its metric
$g_{\alpha\beta}(x)$ and by its metric-compatible connection
$\Gamma_{\alpha\beta}^\mu$, which is used to define the covariant derivative
of
a vector as
\begin{equation}
D_\nu A^\mu = \partial_\nu A^\mu + \Gamma_{\nu\rho}^\mu A^\rho.
\label{cov}
\end{equation}
The $U_4$ connection is non-symmetric in its lower indices, and from its
anti-symmetric part can be defined the torsion tensor
\begin{equation}
S_{\alpha\beta}^{\ \ \gamma} = \frac{1}{2}
\left(\Gamma_{\alpha\beta}^\gamma-\Gamma_{\beta\alpha}^\gamma \right).
\label{torsion}
\end{equation}
One can write the connection as a function of the torsion tensor
\begin{equation}
\Gamma_{\alpha\beta}^\gamma = \left\{_{\alpha\beta}^\gamma \right\}
 + S_{\alpha\beta}^{\ \ \gamma}
- S_{\beta\ \alpha}^{\ \gamma\ } + S_{\ \alpha\beta}^{\gamma\ \ },
\label{connection}
\end{equation}
where $\left\{_{\alpha\beta}^\gamma \right\}$ are
 the usual Christoffel symbols
from Riemannian space-time $V_4$. A quantity that will be particularly useful
is the trace of the connection (\ref{connection}), and using properties of
Christoffel symbols we get the following expression for it
\begin{equation}
\Gamma_{\alpha\beta}^\alpha = \frac{1}{\sqrt{-g}}\partial_\beta \sqrt{-g}
+ 2 S_\beta,
\label{tracegamma}
\end{equation}
where $g$ is the determinant of the metric tensor, and
$S_\beta$ is the trace of
the torsion tensor $S_\beta = S^{\ \ \alpha}_{\alpha\beta}$.

The case where the trace $S_\beta$ can be obtained from a scalar potential
\begin{equation}
S_\beta(x) = \partial_\beta \Theta(x),
\label{pot}
\end{equation}
 will be crucial   in our discussion.
Under the condition (\ref{pot}) we
have the following expression for (\ref{tracegamma})
\begin{equation}
\Gamma_{\alpha\beta}^\alpha =
\frac{e^{-2\Theta}}{\sqrt{-g}}\partial_\beta e^{2\Theta}\sqrt{-g} .
\end{equation}

It is important to note that the often used $V_4$ relation between the
exterior
derivative of an $1$-form and the covariant derivative
\begin{equation}
d A = \partial_\alpha A_\beta dx^\alpha \wedge dx^\beta =
D_\alpha A_\beta dx^\alpha \wedge dx^\beta ,
\label{v4}
\end{equation}
is not valid in $U_4$, where instead of (\ref{v4}) we have\cite{lovelock}
\begin{equation}
d A = \partial_\alpha A_\beta dx^\alpha \wedge dx^\beta =
\left(
D_\alpha A_\beta + \frac{1}{2} S_{\alpha\beta}^{\ \ \rho}A_\rho
\right)dx^\alpha \wedge dx^\beta \neq
D_\alpha A_\beta dx^\alpha \wedge dx^\beta .
\label{u4}
\end{equation}
We will see that the difference between (\ref{v4}) and
(\ref{u4})
is the origin of the problems with the use of MCP in the tensorial
equations of motion in $U_4$.

\section{Abelian fields}

It is well known that Maxwell's equations can be expressed by means of
differential forms and exterior calculus. This description is the most
``economical'', in the sense that it requires the minimal from the
geometry of the manifold. Differential forms and their exterior derivatives
are covariant objects in any differentiable manifold, in spite of  the
manifold is endowed with a connection or not. We will see that this
description can be considered as the most fundamental one, not due
to aesthetic arguments, but by  physical reasons.

In order to study Maxwell's equations in a metric differentiable manifold,
we introduce a fundamental quantity, the electromagnetic potential $1$-form
\begin{equation}
A=A_\alpha dx^\alpha,
\label{potvet}
\end{equation}
and from the potential $1$-form we can define the Faraday's $2$-form
\begin{equation}
F = dA = \frac{1}{2} F_{\alpha\beta}\, dx^\alpha\wedge dx^\beta,
\label{faraday}
\end{equation}
where $F_{\alpha\beta}=\partial_\alpha A_\beta- \partial_\beta A_\alpha$ is
the
usual electromagnetic tensor.

It should be noted that (\ref{potvet}) plays the role of a connection in the
principal bundle ${\cal P}({\cal M},U(1))$, where the base space
${\cal M}$ is the space-time and the electromagnetic gauge group $U(1)$ is the
fiber. If the bundle ${\cal P}({\cal M},U(1))$ is trivial, as for example for
${\cal M}= R^4$, we can assure that a  single gauge connection
(\ref{potvet}) is defined globally.
However, for a non-trivial bundle we can define only locally
the gauge connection. This is  the Dirac monopole case where, due
to the ${\cal P}(S^2,U(1))$ non-triviality, we need at least two gauge
potentials to describe it \cite{nakahara}.
We will ignore by now these problems.

The homogenous Maxwell's equations arise naturally due to the definition
(\ref{faraday}) as a consequence of Poincar\'e's lemma \cite{lovelock}
\begin{equation}
dF = d(dA) = \frac{1}{2}\partial_\gamma F_{\alpha\beta}\,
dx^\gamma\wedge dx^\alpha \wedge dx^\beta = 0,
\label{1st}
\end{equation}
and in terms of components we have
\begin{equation}
\partial_{[\gamma} F_{\alpha\beta]} = 0,
\label{1stc}
\end{equation}
where ${}_{[\ \ \ ]}$ means antisymmetrization.

The non-homogenous equations in Minkowski space-time are given by
\begin{equation}
d {}^*\!F = 4\pi {}^*\! J,
\label{2nd}
\end{equation}
where ${}^*\!J = \frac{1}{3!} \varepsilon_{\alpha\beta\gamma\delta}J^\alpha
dx^\beta\wedge dx^\gamma \wedge dx^\delta$ is the current
$3$-form constructed from the current vector $J^\delta$ and
\begin{equation}
{}^*\!F = \frac{1}{4}\varepsilon_{\alpha\beta\gamma\delta} F^{\gamma\delta}\,
dx^\alpha \wedge dx^\beta,
\label{fstar}
\end{equation}
is the dual of Faraday's $2$-form, constructed from it by using
$\varepsilon_{\alpha\beta\gamma\delta}$,  the totally antisymmetric
symbol, and the metric tensor.

By an accurate analysis of (\ref{2nd}) one can see that it is not covariant
in a curved space-time,
because of ${}^*\!F$ is not a scalar $2$-form, but it is a relative scalar
$2$-form with weight $-1$, due to the anti-symmetrical symbol.
Now we assume that the manifold is endowed with a
connection to use it in order to cast (\ref{2nd}) in a covariant way. This is
\newpage
\noindent
done by substituting the exterior derivative by the covariant one
\begin{equation}
d{}^*\!F \rightarrow {\cal D}{}^*\!F = \frac{1}{3!}\left(
\partial_\alpha {}^*\! F_{\beta\gamma} +
\Gamma^\rho_{\rho\alpha}{}^*\! F_{\beta\gamma}
\right)\delta^{\alpha\beta\gamma}_{\mu\nu\omega} \,
dx^\mu \wedge dx^\nu \wedge dx^\omega,
\label{covex}
\end{equation}
where $\delta^{\alpha\beta\gamma}_{\mu\nu\omega}$ is the generalized Kronecker
delta.
The covariant exterior derivative in (\ref{covex})  takes into account
that
${}^*\!F_{\alpha\beta} =
\frac{1}{2}\varepsilon_{\alpha\beta\gamma\delta}F^{\gamma\delta}$
is a relative $(0,2)$ tensor with weight $-1$.
One can check that ${\cal D}{}^*\!F$ is a relative scalar $3$-form with
weight $-1$.
We have then the following
covariant generalization of (\ref{2nd})
\begin{equation}
{\cal D} {}^*\!F = 4\pi {}^*\! J.
\label{2ndcov}
\end{equation}
Equations (\ref{1st}) are already in a covariant form in any differentiable
manifold.

The informations about the geometry of the manifold are contained in the
metric tensor $g_{\alpha\beta}(x)$ used in the construction of
${}^*\!F$ (\ref{fstar}), and in the trace of the connection used to define the
covariant exterior derivative (\ref{covex}). Therefore we can think that
equations (\ref{2ndcov}) and (\ref{1st}) were obtained from Minkowskian
ones by means of MCP used at differential forms level.
The components expression for
(\ref{2ndcov}) in an Riemann-Cartan space is
\begin{equation}
\frac{1}{\sqrt{-g}}\partial_\mu \sqrt{-g} F^{\nu\mu}
+ 2 S_\mu F^{\nu\mu} = 4\pi J^\nu.
\label{2ndcovc}
\end{equation}
One can see that equation (\ref{2ndcovc}), obtained by
using MCP at differential forms level,
 allows the interaction of
electromagnetism with torsion of space-time without destroying gauge
invariance.

Taking the covariant exterior derivative in both sides of (\ref{2ndcov}) we
get
\begin{equation}
4\pi {\cal D} {}^*\! J = \frac{1}{4!}\left(
\partial_\lambda\Gamma^\rho_{\rho\mu} \right) {}^*\! F_{\nu\omega}
\delta^{\lambda\mu\nu\omega}_{\alpha\beta\gamma\delta}\,
dx^\alpha \wedge dx^\beta \wedge dx^\gamma \wedge dx^\delta,
\label{conserv}
\end{equation}
and to have a generalized conservation condition for the current we need that
\begin{equation}
\partial_\lambda \Gamma^\rho_{\rho\mu} -
\partial_\mu \Gamma^\rho_{\rho\lambda} = 0,
\end{equation}
which has, at least locally,  as general solution \cite{lovelock}
\begin{equation}
\Gamma^\rho_{\rho\mu} = \partial_\mu f(x).
\label{condit}
\end{equation}
Using that $\left\{^\rho_{\rho\mu} \right\}=\partial_\mu \ln \sqrt{-g}$,
equation (\ref{condit})
will have general solution only if the trace of the torsion
tensor obeys (\ref{pot}). In this case
$f(x)=\ln \left(e^{2\Theta}\sqrt{-g}\right)$. When $J=0$, the condition
(\ref{condit}) is a consistency condition for equation (\ref{2ndcov}). Under
the condition (\ref{pot}) we have the following components expression for
(\ref{2ndcov})
\begin{equation}
\frac{e^{-2\Theta}}{\sqrt{-g}}\partial_\mu e^{2\Theta}\sqrt{-g} F^{\nu\mu}=
4\pi J^\nu,
\label{2ndcovv}
\end{equation}
and for the generalized conservation condition we have
\begin{equation}
\frac{e^{-2\Theta}}{\sqrt{-g}}\partial_\mu e^{2\Theta}\sqrt{-g} J^\mu = 0.
\label{newcons}
\end{equation}
It must be stressed that if the trace of the torsion tensor does not obey
(\ref{pot}) we cannot obtain a generalized conservation condition.

One can ask now if it is possible to obtain the non-homogeneous equations
(\ref{2ndcovv}) from an Action principle. We know that in Minkowski
space-time,
the non-homogeneous equations are gotten from the following action
\begin{equation}
S = -\int \left(4\pi{}^*\!J\wedge A +\frac{1}{2} F \wedge {}^*\!F\right).
\label{actmink}
\end{equation}
Besides the metrical tensor, the
 unique non-covariant term in (\ref{actmink}) is the implicit measure
\begin{equation}
dv = \frac{1}{4!}\varepsilon_{\alpha\beta\gamma\delta}
dx^\alpha \wedge dx^\beta \wedge dx^\gamma \wedge dx^\delta.
\end{equation}
 In order to
get a covariant measure we need to introduce a scalar density. In this case
the choice
\begin{equation}
dv=\frac{1}{4!}e^{2\Theta}\sqrt{-g}\varepsilon_{\alpha\beta\gamma\delta}
dx^\alpha \wedge dx^\beta \wedge dx^\gamma \wedge dx^\delta,
\label{dv}
\end{equation}
 leads to
a covariantly constant measure\cite{saa},
$D_\mu e^{2\Theta}\sqrt{-g}=0$. With this new measure one gets the
following coordinate expression for (\ref{actmink})
\begin{equation}
S = \int d^4x \, e^{2\Theta} \sqrt{-g}\left(
-\frac{1}{4}F_{\alpha\beta} F^{\alpha\beta} + 4\pi J^\alpha A_\alpha
\right).
\label{actu4}
\end{equation}
It is easy to check that we can obtain equations (\ref{2ndcovv}) from the
action (\ref{actu4}). We can check also that equations (\ref{1stc}) and
(\ref{2ndcovv}) are invariant under the usual $U(1)$ gauge transformation
\begin{equation}
A_\mu \rightarrow A_\mu + \partial_\mu \varphi.
\label{u1gauge}
\end{equation}
We would like to stress the importance of the generalized conservation
condition (\ref{newcons}) to guarantee the gauge invariance of the
action (\ref{actu4}).

Now we can discuss the usual
 way of coupling Maxwell fields to torsion, in which
 one applies MCP at the level of tensorial equations.
In the usual way of coupling, we loose gauge invariance and also the
homogeneous equation (\ref{1stc}).
Applying usual MCP to the
equation (\ref{1stc}) we get
\begin{equation}
\partial_{[\alpha} \tilde{F}_{\beta\gamma]} +
2 S_{[\alpha\beta}^{\ \ \rho} \tilde{F}_{\gamma]\rho} = 0 ,
\label{falac}
\end{equation}
where $\tilde{F}_{\alpha\beta} =
F_{\alpha\beta} - 2 S_{\alpha\beta}^{\ \ \rho}A_\rho$. One can check
that (\ref{falac}) has no general solutions.
The origin of the problem is the difference between exterior derivatives
and covariant ones
pointed out in section 2. In $V_4$ there exist an ``equivalence''
between exterior derivatives and covariant ones, and due to this in $V_4$
it makes no difference if one applies MCP at differential forms or at
tensorial levels. In $U_4$ we have another situation, and due to the
``inequivalence'' between exterior and covariant derivatives we don't get
the same result applying MCP at different levels. Based in these facts
one claims that the differential forms representation for Maxwell equations
are the most fundamental one.

\section{Non-Abelian fields}

In order to generalize the results of section 3 for the non-abelian case one
needs to introduce the  non-abelian potential $1$-form
\begin{equation}
A = A_\mu^a \lambda ^a dx^\mu,
\label{npot}
\end{equation}
where $\lambda^a$ are the generators of the gauge Lie group ${\cal G}$,
\begin{equation}
\left[ \lambda^a,\lambda^b\right] = f^{abc} \lambda^c.
\end{equation}
Latin indices are reserved to the group manifold, and the summation convention
for repeated indices is adopted. Let us restrict ourselves to compact
semisimple
Lie groups, so that the structure constants are anti-symmetrical under the
change of any couple of indices. Any element $g \in {\cal G}$ can be
written as
\begin{equation}
g(x) = \exp i \theta^a(x) \lambda^a,
\end{equation}
where $\theta^a(x)$ are the group continuous parameters.

Here it's important the same comment already made in Section 2. The gauge
potential $1$-form (\ref{npot}) plays the role of a connection in the
principal
bundle ${\cal P}({\cal M},{\cal G})$, and we can assure the global
validity of a single gauge potential only for trivial bundles. \cite{nakahara}

 From (\ref{npot}) we can
define the $2$-form equivalent to (\ref{faraday}),
\begin{equation}
F =DA=dA + A\wedge A = \frac{1}{2}F^a_{\mu\nu} \lambda^a dx^\mu\wedge dx^\nu,
\end{equation}
where $F^a_{\mu\nu}=\partial_\mu A_\nu^a - \partial_\nu A_\mu^a +
f^{abc}A^b_\mu A^c_\nu$  is the usual non-abelian strength tensor. The
derivative $D$ is the covariant derivative that has the appropriated
transformation law under gauge transformations.

As in the abelian case, the homogeneous  non-abelian equations are a
consequence of Bianchi identity
\begin{equation}
DF = dF + A\wedge F - F\wedge A = \frac{1}{2}
\left(
\partial_\mu F^a_{\omega\nu} + A^b_\mu F^c_{\omega\nu}f^{abc}
\right)\lambda^a dx^\mu\wedge dx^\omega \wedge dx^\nu = 0.
\label{hom}
\end{equation}
The non-homogenous equation for non-abelian gauge fields
are written as the Maxwell ones
(\ref{2nd}). For simplicity and without generality loss, we will treat the
case without sources:
\begin{equation}
D{}^*\!F = d{}^*\!F + A\wedge {}^*\!F - {}^*\!F\wedge A = \frac{1}{2}
\left(
\partial_\mu {}^*\!F^a_{\omega\nu} + A^b_\mu {}^*\!F^c_{\omega\nu}f^{abc}
\right)\lambda^a dx^\mu\wedge dx^\omega \wedge dx^\nu = 0,
\label{nhom}
\end{equation}
where
the dual of the non-abelian strength tensor is defined as in (\ref{fstar}).
In the same way of abelian case, equation (\ref{hom}) is already in a
covariant form, but due to the term ${}^*\!F$, equation (\ref{nhom}) must be
generalized in a curved space-time. To cast (\ref{nhom}) in a covariant way,
one needs to substitute
$d{}^*\!F\rightarrow {\cal D}{}^*\!F$ in (\ref{nhom}) as we did
in (\ref{covex}).
The derivative ${\cal D}$ is defined as
\begin{equation}
{\cal D} {}^*\!F = d {}^*\! F + \omega \wedge {}^*\!F,
\label{covex1}
\end{equation}
where $\omega = \Gamma^\rho_{\rho\alpha} dx^\alpha $. We can check that
(\ref{covex1}) is equivalent to (\ref{covex}). Using the derivative
${\cal D}$ we get the following generalization for (\ref{nhom})
\begin{eqnarray}
D{}^*\!F &=& d{}^*\!F + \omega \wedge {}^*\! F +
 A\wedge {}^*\!F - {}^*\!F\wedge A = \nonumber \\
&=& \frac{1}{2}
\left(
\partial_\mu {}^*\!F^a_{\omega\nu} +
\Gamma^\rho_{\rho\mu}{}^*\!F^a_{\omega\nu} +
A^b_\mu {}^*\!F^c_{\omega\nu}f^{abc}
\right)\lambda^a dx^\mu\wedge dx^\omega \wedge dx^\nu = 0.
\label{nhom1}
\end{eqnarray}
In  order to equations (\ref{hom}) and (\ref{nhom1}) have
non trivial solutions one needs that $D(DF) = D(D{}^*\!F) =0$. Using the
fact that $DF$ and $D{}^*\!F$ are respectively an $3$-form and a
relative $3$-form with weight $-1$, we can obtain
\begin{equation}
D(DF) = d(DF) + A\wedge (DF) + (DF)\wedge A = 0,
\end{equation}
for the homogenous equation (\ref{hom}). For the case of the
non-homogeneous equation (\ref{nhom1}) we have
\begin{equation}
D(D{}^*\!F) = d(D{}^*\!F) + \omega \wedge D{}^*\!F
+ A\wedge (D{}^*\!F) + (D{}^*\!F)\wedge A
= d\omega \wedge {}^*\!F,
\end{equation}
and to get the desired condition $D(D{}^*\!F)=0$, we are enforced to have
 $d\omega=0$.  Since $\omega$  is an $1$-form and it is closed, by the
converse
of Poincar\'e lemma, we have at least in a star-shaped region that
$\omega$ is exact, $\omega = df$, what is the same result that we got in
the abelian case.

Under the hypothesis (\ref{pot}) we have the usual coordinate expression
for (\ref{hom})
\begin{equation}
\varepsilon^{\alpha\beta\gamma\delta}\left(
\partial_\beta F^a_{\gamma\delta} + A^b_\beta F^c_{\gamma\delta}f^{abc}
\right) = 0,
\label{gauge1}
\end{equation}
and the following expression for the generalized non-homogeneous equation
(\ref{nhom})
\begin{equation}
\frac{e^{-2\Theta}}{\sqrt{-g}} \partial_\mu e^{2\Theta}\sqrt{-g}
F^{a\,\nu\mu} + A^b_\mu F^{c\,\nu\mu}f^{abc} = 0.
\label{gauge2}
\end{equation}
One can check that the equations (\ref{gauge1}) and (\ref{gauge2}) are
invariant under the usual non-abelian gauge transformation
\begin{equation}
A_\mu \rightarrow g A_\mu g^{-1} + g \partial_\mu g^{-1} ,
\label{naga}
\end{equation}
where $A_\mu = A^a_\mu \lambda^a$. It is clear from (\ref{gauge2}) that
non-abelian gauge fields are sensitive to the non-Riemannian structure of
space-time.

As in the abelian case, one can try to get equation (\ref{gauge2}) from
an Action principle. We know that in Minkowski space-time the
non-homogeneous equations are gotten form the action
\begin{equation}
S=-\frac{1}{2}\int {\rm trace}\left( F\wedge {}^*\!F \right),
\label{a2}
\end{equation}
which has the following coordinate expression
\begin{equation}
S = -\frac{1}{4} \int d^4 x\, {\rm trace} \left(
F^a_{\mu\nu} F^{b\,\mu\nu} \lambda^a \lambda^b
\right) =
-\frac{1}{4} \int d^4 x F^a_{\mu\nu} F^{a\,\mu\nu},
\end{equation}
where the normalization condition:
${\rm trace}(\lambda^a \lambda^b)=\delta^{ab}$, was assumed for the group
generators.

In order to cast (\ref{a2}) in a covariant way, one needs to substitute the
metric tensor used in the definition of the dual and to modify the measure
of integration. We pick the same measure used in the abelian case
(\ref{dv}), and get
\begin{equation}
S = -\frac{1}{2}\int  e^{2\Theta} \sqrt{-g} \,
{\rm trace} (F \wedge {}^*\!F),
\end{equation}
which has the following coordinate expression
\begin{equation}
S = -\frac{1}{4} \int d^4x \, e^{2\Theta}\sqrt{-g}   F^a_{\mu\nu}
F^{a\,\mu\nu}.
\label{naa}
\end{equation}
Equations (\ref{gauge2}) follow from minimization of (\ref{naa}). It's easy
to realize that the action (\ref{naa}) is
 invariant under non-abelian gauge transformation
(\ref{naga}).

\section{Final Considerations}

Let us summarize the results of previous sections. The Action for gauge
fields in $U_4$ space time is given by
$$
S = -\frac{1}{2}\int  e^{2\Theta} \sqrt{-g} \,
{\rm trace} (F \wedge {}^*\!F)
 = -\frac{1}{4} \int d^4x \, e^{2\Theta}\sqrt{-g}   F^a_{\mu\nu} F^{a\,\mu\nu}.
$$
The $U_4$ equations of motion are
\begin{eqnarray}
\varepsilon^{\alpha\beta\gamma\delta}\left(
\partial_\beta F^a_{\gamma\delta} + A^b_\beta F^c_{\gamma\delta}f^{abc}
\right) &=& 0, \nonumber \\
\frac{e^{-2\Theta}}{\sqrt{-g}} \partial_\mu e^{2\Theta}\sqrt{-g}
F^{a\,\nu\mu} + A^b_\mu F^{c\,\nu\mu}f^{abc} &=& 0, \nonumber
\end{eqnarray}
where the last equations follow from the Action principle. It's easy to check
that the abelian limit of these results corresponds to the Maxwell model of
Section 3. These results are valid in an Riemann-Cartan space-time where the
trace of the torsion tensor can be derived from a scalar potential,
$S_\alpha = \partial_\alpha\Theta$. On the other hand, we cannot get
consistent equations if the trace cannot be obtained from a scalar potential.

It should be noted that the existence of the scalar $\Theta$, such that
$S_\alpha = \partial_\alpha\Theta$, was assured by the converse of the
Poincar\'e lemma, and then, one cannot assure that only one scalar
$\Theta$ is enough to define globally the trace $S_\alpha$. This will
depend on the topology of the space-time manifold ${\cal M}$. As an
example, for ${\cal M} = R^4$ we can define the trace $S_\alpha$ globally
from an unique scalar $\Theta$.

In our discussion, the space-time manifold is considered as independent
from the gauge fields. Gauge fields impose some restrictions on the
geometry, but its dynamics is not affected by the gauge fields. The geometry
is treated  as an
external field. But we know, from General Relativity, that the dynamics
of the space-time geometry must be governed by the non-gravitational fields,
in this case the gauge fields. An interesting point is to introduce the
geometry of the space-time in the discussion, by adding a term for it
in the Action principle.

With the assumption that the dynamics of the
$U_4$ geometry is given by an Einstein-Hilbert action with the volume
element (\ref{dv}), we get the following action for the full system:
\begin{equation}
S = - \int d^4x \, e^{2\Theta}\sqrt{-g}\left(
\frac{1}{4}F^a_{\mu\nu} F^{a\,\mu\nu} + R \right),
\label{fullaction}
\end{equation}
where $R$ is the scalar of curvature, calculated from the  connection
(\ref{connection}) with the conventions of the reference \cite{venzo}. From
the variation of (\ref{fullaction}) with respect to $g_{\mu\nu}$ and
$S_{\alpha\beta\gamma}$
one gets the following equations for the $U_4$ geometry:
\begin{eqnarray}
\label{fulleq}
&& R_{\mu\nu}^{V_4} = 2D_\mu S_\nu - \frac{4}{3}g_{\mu\nu}S_\rho S^\rho
-\frac{1}{2} \left(F^a_{\mu\alpha}F^{a\,\ \alpha}_{\ \nu }
+\frac{1}{2}g_{\mu\nu} F^a_{\omega\rho}F^{a\,\omega\rho} \right), \nonumber \\
&& D_\mu S^\mu = -\frac{3}{8}F_{\mu\nu}F^{\mu\nu}, \\
&& \tilde{S}_{\alpha\beta\gamma}=0. \nonumber
\end{eqnarray}
In (\ref{fulleq}),
$\tilde{S}_{\alpha\beta\gamma}$ is the tracelles part of the torsion tensor
\begin{equation}
S_{\alpha\beta\gamma} = \tilde{S}_{\alpha\beta\gamma} + \frac{1}{3}
\left(g_{\alpha\gamma} S_\beta - g_{\beta\gamma} S_\alpha \right),
\end{equation}
and $R^{V_4}_{\mu\nu}$ is the $V_4$ Ricci tensor, calculated from the
Christoffel symbols. The equations (\ref{fulleq}) point out new features
of Einstein-Cartan theory of gravity with the new volume element \cite{saa1}.
In particular, one can see that gauge fields are sources of torsion, and that
the torsion propagates.

The similarity between (\ref{fullaction}) and the action for dilaton gravity,
derived from a perturbative approach for bosonic strings
\cite{dilaton,green},
 is surprising. The peculiar $\Theta$-dependence of the measure in
(\ref{fullaction}) is identical to the dilaton dependence of the measure in the
effective action for gravity with one-loop string-theoretic  corrections.
 These topics are now under investigation.

The problems of covariance of the equations of motion always was connected
with the duals of the strength tensors, and we would like to dedicate the
last subsection to this topic.

\subsection{Hodge star operator}

 The mathematical essence of the problems with the covariance presented in the
last two sections, is the duality transformations, i.e. Hodge star $({}^*)$
operation\cite{nakahara}. The problems with covariance could be avoided if
one changes appropriately the Hodge star operator for an $U_4$ manifold. For
this purpose, we introduce the ${}^*$ operator following \cite{nakahara}.

Be ${\cal M}$ a $n$-dimensional differentiable manifold endowed with a
metric $g_{\mu\nu}$ and with a metric-compatible connection
$\Gamma^\alpha_{\gamma\beta}$, and $\Omega^m({\cal M})$ the  space of
differential $m$-forms on it. The Hodge ${}^*$ operator is a linear operator
\begin{equation}
{}^* : \Omega^m({\cal M}) \rightarrow  \Omega^{n-m}({\cal M}),
\end{equation}
which for a Riemannian manifold has the following action on a basis vector
of $\Omega^m({\cal M})$
\begin{equation}
{}^* ( dx^{\alpha_1} \wedge dx^{\alpha_2} \wedge ... \wedge dx^{\alpha_m}) =
\frac{\sqrt{g}}{(n-m)!}
\varepsilon^{\alpha_1 ... \alpha_m}_
{\ \ \ \ \ \ \ \beta_{m+1}...\beta_n}
dx^{\beta_{m+1}}\wedge ... \wedge dx^{\beta_n},
\label{hodge}
\end{equation}
where $\varepsilon_{\alpha_1 ... \alpha_n}$ is the totally anti-symmetrical
symbol. The action of (\ref{hodge}) on the basis vector for
$\Omega^0({\cal M})$ gives
\begin{equation}
{}^* 1 = \frac{\sqrt{g}}{n!} \varepsilon_{\alpha_1 ... \alpha_n}
dx^{\alpha_1}\wedge ... \wedge dx^{\alpha_n} = \sqrt{g} d^4 x,
\label{ele}
\end{equation}
that is the invariant and covariantly constant volume element for a Riemannian
manifold.

In an Riemann-Cartan space-time, the volume element (\ref{ele}) is not
covariantly constant, in contrast to the Riemannian case, as one can check
using the fact that $\sqrt{g}$ is a scalar density
\begin{equation}
D_\mu\sqrt{g} = \partial_\mu\sqrt{g} - \Gamma^\alpha_{\alpha\mu}\sqrt{g} =
-2S_\mu\sqrt{g}.
\end{equation}
 To get an
invariant and covariantly constant volume element for an Riemann-Cartan
space-time that obeys (\ref{pot}) one needs to modify the Hodge ${}^*$
operator by
\begin{equation}
{}^* ( dx^{\alpha_1} \wedge dx^{\alpha_1} \wedge ... \wedge dx^{\alpha_1}) =
\frac{h(x)}{(n-m)!}
\varepsilon^{\alpha_1 ... \alpha_m}_
{\ \ \ \ \ \ \ \beta_{m+1}...\beta_n}
dx^{\beta_{m+1}}\wedge ... \wedge dx^{\beta_n},
\label{nhodge}
\end{equation}
where the scalar density $h(x)$ is such that
$\partial_\mu h = \Gamma_{\nu\mu}^\nu h$. We already know that
$h(x) = e^{2\Theta}\sqrt{g}$. Using (\ref{nhodge}) to define
the duals used in  the equations of motion and in the Lagrangian we will get
automatically the $U_4$ covariant equations.

It is not clear if one can define a Hodge ${}^*$ operator in order to obtain
a invariant and covariantly constant volume element for the case of
Riemann-Cartan space-times not obeying (\ref{pot}).

\acknowledgements

The author is grateful to Professors Josif Frenkel and \'Elcio Abdalla for
discussions, and to Funda\c c\~ao
de Amparo \`a Pesquisa do Estado de S\~ao Paulo for support.

\end{document}